\input harvmac
\noblackbox
\newcount\figno
\figno=0
\def\fig#1#2#3{
\par\begingroup\parindent=0pt\leftskip=1cm\rightskip=1cm\parindent=0pt
\baselineskip=11pt
\global\advance\figno by 1
\midinsert
\epsfxsize=#3
\centerline{\epsfbox{#2}}
\vskip 12pt
\centerline{{\bf Figure \the\figno:} #1}\par
\endinsert\endgroup\par}
\def\figlabel#1{\xdef#1{\the\figno}}

\def\np#1#2#3{Nucl. Phys. {\bf B#1} (#2) #3}
\def\pl#1#2#3{Phys. Lett. {\bf B#1} (#2) #3}
\def\prl#1#2#3{Phys. Rev. Lett. {\bf #1} (#2) #3}
\def\prd#1#2#3{Phys. Rev. {\bf D#1} (#2) #3}


\font\cmss=cmss10
\font\cmsss=cmss10 at 7pt
\def\rlx{\relax\leavevmode}
\def\inbar{\vrule height1.5ex width.4pt depth0pt}
\def\IC{\relax\,\hbox{$\inbar\kern-.3em{\rm C}$}}
\def\IN{\relax{\rm I\kern-.18em N}}
\def\IP{\relax{\rm I\kern-.18em P}}
\def\ZZ{\rlx\leavevmode\ifmmode\mathchoice{\hbox{\cmss Z\kern-.4em Z}}
 {\hbox{\cmss Z\kern-.4em Z}}{\lower.9pt\hbox{\cmsss Z\kern-.36em Z}}
 {\lower1.2pt\hbox{\cmsss Z\kern-.36em Z}}\else{\cmss Z\kern-.4em
 Z}\fi}
\def\IZ{\relax\ifmmode\mathchoice
{\hbox{\cmss Z\kern-.4em Z}}{\hbox{\cmss Z\kern-.4em Z}}
{\lower.9pt\hbox{\cmsss Z\kern-.4em Z}}
{\lower1.2pt\hbox{\cmsss Z\kern-.4em Z}}\else{\cmss Z\kern-.4em
Z}\fi}

\def\narrowplus{\kern -.04truein + \kern -.03truein}
\def\narrowminus{- \kern -.04truein}
\def\narrowminussub{\kern -.02truein - \kern -.01truein}

\def\g{{\gamma}}
\def\e{{\epsilon}}

\def\r{{\rightarrow}}

\def\frac#1#2{{#1\over #2}}

\def\IZ{\relax\ifmmode\mathchoice
{\hbox{\cmss Z\kern-.4em Z}}{\hbox{\cmss Z\kern-.4em Z}}
{\lower.9pt\hbox{\cmsss Z\kern-.4em Z}}
{\lower1.2pt\hbox{\cmsss Z\kern-.4em Z}}\else{\cmss Z\kern-.4em
Z}\fi}
\def\IB{\relax{\rm I\kern-.18em B}}
\def\IC{{\relax\hbox{$\inbar\kern-.3em{\rm C}$}}}
\def\ID{\relax{\rm I\kern-.18em D}}
\def\IE{\relax{\rm I\kern-.18em E}}
\def\IF{\relax{\rm I\kern-.18em F}}
\def\IG{\relax\hbox{$\inbar\kern-.3em{\rm G}$}}
\def\IGa{\relax\hbox{${\rm I}\kern-.18em\Gamma$}}
\def\IH{\relax{\rm I\kern-.18em H}}
\def\II{\relax{\rm I\kern-.18em I}}
\def\IK{\relax{\rm I\kern-.18em K}}
\def\IP{\relax{\rm I\kern-.18em P}}

\font\cmss=cmss10 \font\cmsss=cmss10 at 7pt
\def\IR{\relax{\rm I\kern-.18em R}}

\def\f{\psi}

\def\dd{ {\delta}}

\def\df{\dot{\psi}}

\def\RV{{R_\Vert}}
%

%
%
\def\eqnn#1{\xdef #1{(\secsym\the\meqno)}\writedef{#1\leftbracket#1}%
\global\advance\meqno by1\wrlabeL#1}
\def\eqna#1{\xdef #1##1{\hbox{$(\secsym\the\meqno##1)$}}
\writedef{#1\numbersign1\leftbracket#1{\numbersign1}}%
\global\advance\meqno by1\wrlabeL{#1$\{\}$}}
\def\eqn#1#2{\xdef #1{(\secsym\the\meqno)}\writedef{#1\leftbracket#1}%
\global\advance\meqno by1$$#2\eqno#1\eqlabeL#1$$}



\lref\rK{N. Ishibashi, H. Kawai, Y. Kitazawa and A. Tsuchiya, hep-th/9612115.}
\lref\rCallias{C. Callias, Commun. Math. Phys. {\bf 62} (1978), 213.}
\lref\rPD{J. Polchinski, hep-th/9510017, \prl{\bf 75}{1995}{47}.}
\lref\rWDB{E. Witten,  hep-th/9510135, Nucl. Phys. {\bf B460} (1996) 335.}
\lref\rSSZ{S. Sethi, M. Stern, and E. Zaslow, Nucl. Phys. {\bf B457} (1995)
484.}
\lref\rGH{J. Gauntlett and J. Harvey, Nucl. Phys. {\bf B463} 287. }
\lref\rAS{A. Sen, Phys. Rev. {\bf D53} (1996) 2874; Phys. Rev. {\bf D54} (1996)
2964.}
\lref\rWI{E. Witten, Nucl. Phys. {\bf B202} (1982) 253.}
\lref\rPKT{P. K. Townsend, Phys. Lett. {\bf B350} (1995) 184.}
\lref\rWSD{E. Witten, Nucl. Phys. {\bf B443} (1995) 85.}
\lref\rASS{A. Strominger, Nucl. Phys. {\bf B451} (1995) 96.}
\lref\rBSV{M. Bershadsky, V. Sadov, and C. Vafa, Nucl. Phys. {\bf B463}
(1996) 420.}
\lref\rBSS{L. Brink, J. H. Schwarz and J. Scherk, Nucl. Phys. {\bf B121}
(1977) 77.}
\lref\rCH{M. Claudson and M. Halpern, Nucl. Phys. {\bf B250} (1985) 689.}
\lref\rSM{B. Simon, Ann. Phys. {\bf 146} (1983), 209.}
\lref\rGJ{J. Glimm and A. Jaffe, {\sl Quantum Physics, A Functional Integral
Point of View},
Springer-Verlag (New York), 1981.}
\lref\rADD{ U. H. Danielsson, G. Ferretti, B. Sundborg, Int. J. Mod. Phys. {\bf
A11} (1996) 5463\semi   D. Kabat and P. Pouliot, Phys. Rev. Lett. {\bf 77}
(1996), 1004.}
\lref\rDKPS{ M. R. Douglas, D. Kabat, P. Pouliot and S. Shenker,
hep-th/9608024,
Nucl. Phys. {\bf B485} (1997), 85.}
\lref\rhmon{S. Sethi and M. Stern, Phys. Lett. {\bf B398} (1997), 47.}
\lref\rBFSS{T. Banks, W. Fischler, S. H. Shenker, and L. Susskind,
Phys. Rev. {\bf D55} (1997) 5112.}
\lref\rBHN{ B. de Wit, J. Hoppe and H. Nicolai, Nucl. Phys. {\bf B305}
(1988), 545\semi
B. de Wit, M. M. Luscher, and H. Nicolai, Nucl. Phys. {\bf B320} (1989),
135\semi
B. de Wit, V. Marquard, and H. Nicolai, Comm. Math. Phys. {\bf 128} (1990),
39.}
\lref\rT{ P. Townsend, Phys. Lett. {\bf B373} (1996) 68.}
\lref\rLS{L. Susskind, hep-th/9704080.}
\lref\rFH{J. Frohlich and J. Hoppe, hep-th/9701119.}
\lref\rAg{S. Agmon, {\it Lectures on Exponential Decay of Solutions of
Second-Order Elliptic Equations}, Princeton University Press (Princeton) 1982.}
\lref\rY{P. Yi, hep-th/9704098.}
\lref\rDLhet{ D. Lowe, hep-th/9704041.}
\lref\rqm{M. Claudson and M. Halpern, \np{250}{1985}{689}\semi
R. Flume, Ann. Phys. {\bf 164} (1985) 189\semi
M. Baake, P. Reinecke and V. Rittenberg, J. Math. Phys. {\bf 26} (1985) 1070.}
\lref\rbb{K. Becker and M. Becker, hep-th/9705091, \np{506}{1997}{48}\semi
K. Becker, M. Becker, J. Polchinski and A. Tseytlin, hep-th/9706072,
\prd{56}{1997}{3174}.}
\lref\rss{S. Sethi and M. Stern, hep-th/9705046. }
\lref\rpw{J. Plefka and A. Waldron, hep-th/9710104, \np{512}{1998}{460}.}
\lref\rhs{M. Halpern and C. Schwartz, hep-th/9712133.}
\lref\rlimit{N. Seiberg hep-th/9710009, \prl{79}{1997}{3577}\semi
A. Sen, hep-th/9709220.}
\lref\rentin{D.-E. Diaconescu and R. Entin, hep-th/9706059,
\prd{56}{1997}{8045}.}
\lref\rgreen{M. B. Green and M. Gutperle, hep-th/9701093, \np{498}{1997}{195}.}
\lref\rpioline{B. Pioline, hep-th/9804023.}
\lref\rgl{O. Ganor and L. Motl, hep-th/9803108.}
\lref\rds{M. Dine and N. Seiberg, hep-th/9705057, \pl{409}{1997}{209}.}
\lref\rberg{E. Bergshoeff, M. Rakowski and E. Sezgin, \pl{185}{1987}{371}.}
\lref\rBHP{M. Barrio, R. Helling and G. Polhemus, hep-th/9801189.}
\lref\rper{P. Berglund and D. Minic, hep-th/9708063, \pl{415}{1997}{122}.}
\lref\rspin{P. Kraus, hep-th/9709199, \pl{419}{1998}{73}\semi
J. Harvey, hep-th/9706039\semi
J. Morales, C. Scrucca and M. Serone, hep-th/9709063, \pl{417}{1998}{233}.}
\lref\rdine{M. Dine, R. Echols and J. Gray, hep-th/9805007.}
\lref\rber{D. Berenstein and R. Corrado, hep-th/9702108, \pl{406}{1997}{37}.}
\lref\rnonpert{A. Sen, hep-th/9402032, \pl{329}{1994}{217}; hep-th/9402002,
Int. J. Mod. Phys. {\bf
A9} (1994) 3707\semi
N. Seiberg and E. Witten, hep-th/9408099, \np{431}{1995}{484}.
}
\lref\rpss{S. Paban, S. Sethi and M. Stern, hep-th/9805018.}
\lref\rperiwal{V. Periwal and R. von Unge, hep-th/9801121.}
\lref\rfer{M. Fabbrichesi, G. Ferretti and R. Iengo, hep-th/9806018.}
\lref\rdr{ M. Dine and A. Rajaraman, hep-th/9710174.}
\lref\rmet{R. Metsaev, M. Rahmanov and A. Tseytlin, \pl{193}{1987}{207}.}
\lref\rts{I. Chepelev and A. Tseytlin, hep-th/9709087, \np{515}{1998}{73}.}
\lref\rkraus{E. Keski-Vakkuri and P. Kraus, hep-th/9709122, \np{518}{1998}{212}.}
\lref\ralwis{S. de Alwis, hep-th/9710219, \pl{423}{1998}{59}. }

\Title{\vbox{\hbox{hep-th/9806028}
\hbox{DUK-CGTP-98-05, IASSNS--HEP--98/48, UTTG-08-98}}}
{\vbox{\centerline{Supersymmetry and Higher Derivative Terms in the}
\vskip8pt\centerline{Effective Action of Yang-Mills Theories}}}
\centerline{Sonia Paban$^\ast$\footnote{$^1$} {paban@sns.ias.edu}, Savdeep
Sethi$^\spadesuit$\footnote{$^2$} {sethi@sns.ias.edu} and Mark
Stern$^\dagger$\footnote{$^3$} {stern@math.duke.edu} }
\medskip\centerline{ $\ast$ \it Theory Group, Department of Physics,
University of Texas, Austin, TX 78712, USA}
\medskip\centerline{$\spadesuit$ \it School of Natural Sciences, Institute for
Advanced Study, Princeton, NJ 08540, USA}
\medskip\centerline{$\dagger$ \it Department of Mathematics, Duke University,
Durham, NC 27706, USA}

\vskip 0.5in

Higher derivative terms in the effective action of certain Yang-Mills theories
can be severely constrained by supersymmetry. We show that requiring sixteen
supersymmetries in quantum mechanical gauge theory determines the $v^6$ term in
the effective action. Even the numerical coefficient of the $v^6$ term is fixed
in
terms of lower derivative terms in the effective action.

\vskip 0.1in
\Date{6/98}

\newsec{Introduction}

A better understanding of Yang-Mills theories with extended supersymmetry is
crucial
if we are to gain a deeper understanding of the various non-perturbative field
theory
dualities. For example, extended supersymmetry plays a key role in making
possible
conjectures about exact
strong-weak coupling dualities in four-dimensional theories with eight and
sixteen
supersymmetries \rnonpert.
Yang-Mills theories with extended supersymmetry have also played a prominent
role
in a recent attempt to define M theory \rBFSS. In that
endeavor, the theory of interest is the  quantum mechanical gauge theory that
 describes the low-energy dynamics of zero-branes in type IIA string theory
\refs{\rPD, \rWDB}. The system can be obtained by a
 dimensional reduction of supersymmetric Yang-Mills from ten dimensions \rqm.
The theory has sixteen supersymmetries and a $U(N)$ gauge symmetry. For finite
$N$,
this matrix model is believed to describe M theory quantized in the discrete
light-cone
formalism (DLCQ) \refs{\rLS, \rlimit}.

More generally, we should ask the question: to what extent does supersymmetry
determine
the form of the effective action of Yang-Mills theories? In a recent paper, we
proved a
non-renormalization theorem for the $v^4$ term in the effective action of
D0-brane
quantum mechanics \rpss.
The aim of this letter is to apply the same technique to the $v^6$ term to show
that it
is also determined by supersymmetry. Quantum mechanical gauge theory with
sixteen
supersymmetries is a quite subtle theory. Since the coupling has positive mass
dimension, the theory
is strongly coupled at low energies. For example, in matrix theory $g^2 =
M_{pl}^6 \RV^3$
where $\RV$ is the size of the longitudinal direction and $g^2$ is the
Yang-Mills
coupling.

More importantly, the theory has a highly non-trivial vacuum for any $N$ as
conjectured
in \rWDB\ and proven for $N=2$ in \rss. Studying an effective action obtained
by
perturbing around the trivial vacuum is unlikely to make much sense at higher
orders
in a derivative expansion. It seems much like trying to analyze the long
wavelength
physics of QCD using perturbation theory. Indeed, recent arguments suggest that
at
order $v^8$, the perturbative
derivative expansion breaks down \rdine. What should be surprising is that
perturbative
computations actually gave results that agreed with supergravity for the $v^4$
and $v^6$ terms \refs{\rbb, \rDKPS, \rber}.\foot{Comments about a puzzle \rdr\ for 
the $v^6$ terms in higher rank theories have recently appeared in \rfer.} As we 
shall see, the reason
for such
agreement is essentially
the strong constraints imposed by supersymmetry on the effective action. It
seems
likely that the construction of a complete effective action will first require
developing
a somewhat new perturbation theory for scattering amplitudes. A correct
perturbation
theory, along the lines described in \rss,  must encorporate the non-trivial
vacuum
structure. This is a fascinating problem that will be explored elsewhere. That
supersymmetry determines the $F^4$ and $F^6$ terms in ten-dimensional
Yang-Mills has been shown in \rmet. Comments on the general structure of Yang-Mills
effective actions have appeared in \refs{\rts, \rkraus, \ralwis}.

\newsec{Constraining the Six Derivative Terms}

Ignoring acceleration terms, the bosonic part of the D0-brane effective action
takes the form:

\eqn\action{ \, S = \int{ dt \, \left( \, f_1(r) v^2
+ f_2(r) v^4 + f_3(r) v^6 \ldots \, \right).
}}
A discussion of the Lagrangian for four-dimensional Yang-Mills including
acceleration
terms is given in \rperiwal. For the most part, we shall restrict our
discussion
to the effective action
describing the dynamics of two clusters of D0-branes. The Lagrangian contains
both
bosonic fields $x^i$ as well as fermions $ \f_a$, where $i=1,\ldots, 9$ and
$a=1,\ldots,16$.

The $Spin(9)$ Clifford algebra can be represented
by real symmetric matrices $\g^i_{ab}$, where $i=1,\ldots,9$ and
$a=1,\ldots,16$. These matrices satisfy the relation,
\eqn\clifford{ \{ \g^i, \g^j \} = 2 \delta^{ij}, }
and a complete basis contains $\left\{ I, \g^i, \g^{ij},
\g^{ijk}, \g^{ijkl} \right\}$, where we
define:
\eqn\defs{ \eqalign{ \g^{ij} &= {1\over 2!} ( \g^i \g^j - \g^j \g^i) \cr
\g^{ijk} &= {1\over 3!}( \g^i \g^j \g^k - \g^j \g^i \g^k + \ldots) \cr
\g^{ijkl} &= {1\over 4!}( \g^i \g^j \g^k \g^l - \g^j \g^i \g^k \g^l + \ldots).
 \cr}}
The basis decomposes into symmetric, $\left\{ I, \g^i, \g^{ijkl} \right\}$,
and antisymmetric matrices, $\left\{ \g^{ij}, \g^{ijk} \right\}$. The
normalizations in
\defs\ are chosen so that the trace of the square
of a basis element is $ \pm 16$.

Supersymmetry demands that $f_1$ be constant and $f_2 = {c_2\over r^7}$ \rpss.
We will
choose $f_1={1\over 2}$. The coefficient $c_2$ is determined by a one-loop
computation
\rbb. The Lagrangian $L$ can be expressed as the sum of terms,
$L=\sum L_k$, where $L_k$ contains all terms of order $2k$. For example,
\eqn\vsquare{ L_1= \int \, dt \left( {1 \over 2} \, v^2 + i \, \f  \df
\right).}
The order
counts the number of time derivatives plus twice the number of fermions.
Schematically at
order 6,  we
need to consider all terms,

\eqn\vsix{ L_3= \int \, dt \left( f_3^{(0)}(r) \, v^6 + \ldots +
f_3^{(12)}(r) \,\f^{12}
\right), }
which are in the supersymmetric completion of $v^6$. The omitted terms contain
accelerations and fermions with multiple time derivatives. The supersymmetry
transformations take the general form:

\eqn\newtransforms{ \eqalign{ \dd x^i & = -i \e \g^i \f  + \e  N^i \f\cr
\dd \f_a &= ( \g^i v^i \e )_a + ( M \e )_a.}}
The terms $N^i$ and $M$ encode all higher derivative corrections to the
supersymmetry transformations and $\e$ is a sixteen component Grassmann
parameter. Note
that once higher derivative terms appear in
$L$, we must have $N^i$ and $M$ non-zero or the supersymmetry algebra no longer
closes. The actual construction of $N^i$ and $M$ is a tedious business.
Fortunately,
as in the case of the $v^4$ term, we will not need to know very much about
$N^i$ and
$M$ to show that the $v^6$ term is also determined by supersymmetry.

The terms in $L_2$ generate corrections to the supersymmetry transformations of
order
$2$ in $N^i$ and of order 3 in $M$. These corrections are fully determined by
$L_2$.
When we
include the six derivative terms in $L_3$, we get higher derivative terms in
$N^i$ of
order $4$ and in $M$ of order 5. We will only need to know the order of the
terms in
$N^i$ and $M$.

We primarily wish to consider the twelve fermion term which is the `top' form
in the supersymmetric completion of $v^6$. A study of the analogous term in the
completion of $v^4$ gave a non-renormalization theorem for the $v^4$
term.\foot{
It is worth stressing that essentially the same argument used to determine the
four derivative terms in \rpss\ can be applied in any dimension to four
derivative
terms in Yang-Mills theories with only eight supersymmetries and a flat
metric.}
The variation of this term in \vsix\ schematically contains two pieces,

\eqn\variationsix{ \dd ( f_3^{(12)}(r) \f^{12} )  =   \dd f_3^{(12)}(r)  \,\,
\f^{12}
+ f_3^{(12)}(r) \dd  \f^{12}. }
Acting on terms with order 6, we need only consider the lowest order
free-particle
supersymmetry transformations. The variation of $L_3$ then gives terms of order
6,
where we count $\e$ as order $-1/2$. The first term in \variationsix\ contains
a thirteen fermion term. Note that no other term in $L_3$ varies into
the thirteen term. Can any term from $L_1$ vary into a thirteen fermion term?
The
highest order term in $N^i$ is order 4, which can contain an eight fermion
term.
The highest term in $M$ can contain a ten fermion term. It is easy to check
that
the variation of $L_1$ given in \vsquare\ cannot then contain a thirteen
fermion
term.

We can ask the same question about terms from $L_2$. The top form in $L_2$ is
an eight fermion term which is non-vanishing and shown in
\rpss\ to agree with the form computed at one-loop in \rBHP. The relevant term
in $N^i$ is order 2 and so can contain a four fermion term, while the relevant
term in $M$ is order 3 and so can contain a six fermion term. Therefore, a
variation
of the top form in $L_2$ can generate a thirteen fermion term. These are the
only
two sources of thirteen fermion terms in the Lagrangian.

The last piece of information that we need is the number of independent
twelve fermion terms. These terms need to be invariant under the discrete
symmetry which acts as complex conjugation and sends,
$$ x \r -x \qquad \qquad t \r -t. $$
All $n$ fermion structures $T_{a_1\ldots a_n}$ are Hodge dual to $16-n$
fermion structures using the epsilon symbol in sixteen dimensions. We therefore
only need to ask how many independent four fermion structures are possible. It
is easy to check using the Fierz identities in Appendix A of \rpss\ that
the only allowed independent structure is,
\eqn\structure{ x^i x^j \left( \f \g^{ik} \f \f \g^{kj} \f \right).}
Therefore, there is a unique twelve fermion structure,
\eqn\twelve{ T_{a_1\ldots a_{12}} = \e_{a_1\ldots a_{12} b_1 b_2 b_3 b_4} \left
(x^i x^j
 \g^{ik}_{b_1 b_2} \g^{kj}_{b_3 b_4} \right),}
and we define:
$$ T = T_{a_1\ldots a_{12}} \f_{a_1} \cdots \f_{a_{12}}. $$

There are two possible cases: either the terms from $L_2$ make a contribution
to the
thirteen fermion term in the variation of L, or they do not. Let us assume they
do
not make a contribution. This implies that,
\eqn\variation{ \eqalign{ \delta_a \left( f_3^{(12)} T \right)  &= -i \g^s_{ab}
\f_b \, \partial_s \left( f_3^{(12)} \, T \right)\cr & =0.}}
We can apply the operator $ \g^q_{ac}{ d \over {d \f_c}} \partial_q$ to
\variation. After summing over the $a$ index, we learn that,
$$ \Delta \left( f_3^{(12)} T \right) = 0,$$
which gives the equation:
\eqn\close{ {d^2 \over dr^2} f_3^{(12)} + 12{1\over r} { d \over dr} f_3^{(12)}
= 0.}
However, the solution to this equation $f_3^{(12)} \sim 1/r^{11}$ is unphysical
since
it implies that the twelve fermion term
is proportional to a negative power of the coupling. A tree level twelve
fermion term
would need a power of $1/r^{14}$. Actually, we should note that equation
\close\ is
weaker than equation \variation, and one can show directly from \variation\
that the function $f_3^{(12)}$ vanishes.

We should now
consider the case where the terms from $L_2$ do contribute. The terms in $L_2$
are one-loop exact so power counting is easy. The eight fermion term is
proportional
to $1\over r^{11}$. The relevant corrections to the supersymmetry
transformations
have the following dependence on $r$,
$$ \eqalign{ N^i & \sim {1\over r^9} \f^4 \cr M & \sim {1\over r^{10}} \f^6,}
$$
in accord with the one-loop exactness of $L_2$. The equation \close\ then
becomes,
\eqn\final{ {d^2 \over dr^2} f_3^{(12)} + 12{1\over r} { d \over dr} f_3^{(12)}
 + {c_2' \over r^{24}}= 0,}
where the non-zero coefficient $ c_2' $ is determined by the terms in $L_2$.
As we have seen, the homogeneous solution to \final\ where $c_2'=0$ is
unphysical
so we discard
it. The remaining solution gives,
\eqn\solved{ f_3^{(12)} = - {c_2' \over 242} {1\over r^{22}}, }
which is precisely the power needed to agree with a two-loop calculation
like the one  performed in \rbb. Therefore, the $v^6$ terms are also not
renormalized but are
completely determined by supersymmetry and the lower derivative terms in the
effective action. The same method can be used to learn about the six derivative
terms in Yang-Mills theories with sixteen supersymmetries in
various dimensions.

\bigbreak\bigskip\bigskip\centerline{{\bf Acknowledgements}}\nobreak
We would like to thank P. Pouliot for stimulating comments on this topic. The
work of S.S. is supported by NSF grant
DMS--9627351 and that of M.S. by NSF grant DMS-9505040.  S.P. would like to
thank the Institute for Advanced Study for hospitality during the completion of
part of this work.

\listrefs
\bye